# Investigation of transient surface electric field induced by femtosecond laser irradiation of aluminum

Run-Ze Li[1], Pengfei Zhu[1], Long Chen[1], Tong Xu[1], Jie Chen[1,*], Jianming Cao[1,2], Zheng-Ming Sheng[1,3], Jie Zhang[1,*]

[1]*Key Laboratory for Laser Plasmas (Ministry of Education) and Department of Physics and Astronomy, Shanghai Jiao Tong University, Shanghai 200240, China*

[2]*Physics Department and National High Magnetic Field Laboratory, Florida State University, Tallahassee, Florida 32310, USA*

[3]*SUPA, Department of Physics, University of Strathclyde, Glasgow G4 0NG, UK*

[*]Correspondence: Jie Chen: jiec@sjtu.edu.cn; Jie Zhang: jzhang1@sjtu.edu.cn

**Abstract**

Transient surface electric fields induced by femtosecond laser irradiation of an aluminum film were investigated directly by ultrashort electron pulses. At pump intensities of $2.9\sim7.1\times10^{10}$ W/cm$^2$, the transient electric fields last at least one nanosecond with a maximum field strength of $3.2\sim5.3\times10^4$ V/m at 120 μm above the aluminum surface. The transient electric fields and the associated evolution of photoelectrons were explained by a "three-layer" model. The potential influence of such fields on reflection ultrafast electron diffraction and time-resolved angle-resolved photoemission spectroscopy were evaluated.

# 1. Introduction

Transient electric field (TEF) generally exists in femtosecond laser-matter interactions due to thermionic and/or multi-photon emission of electrons [1-5]. The strength and evolution of the TEF is critical in laser ablation mechanism studies [6-8], and the formation of early stage plasmas [9, 10] after intense laser irradiation. Under moderate laser excitation conditions, it is also a nontrivial influencing factor for photocathode optimizations [11-15] and time-resolved electron scattering studies, such as ultrafast electron diffraction (UED) [16-20] and time-resolved angle-resolved photoemission spectroscopy (TR-ARPES) [21-23]. In UED studies, a crystalline sample is first excited by an ultrashort optical pump pulse and then interrogated by an electron probe pulse delivered at a specific delay time. Transient structural information is majorly obtained from the time-dependent evolutions of the diffraction angle and intensity extracted from electron diffraction patterns. However, the existence of the TEF on the sample surface may distort the trajectory of the probe electrons and make the interpretation of diffraction patterns complicated [24-26]. For example, in the studies of semiconductors by reflection UED and metals by transmission UED, the deflection angles induced by the TEFs were comparable with the changes of the diffraction angle originated from structural dynamics [27-29]. In the transmission geometry, it has been demonstrated that the structural dynamics and the TEF effect can be distinguished by simultaneously tracking the radii and the centroids of the diffraction rings of polycrystalline crystals [29]. However, their separation in reflection UED is still indistinct. In such case, the TEF is nearly normal to the propagation direction of the reflective probe electrons, which may induce additional and notable deflections to the probe electrons. Furthermore, the field gradient perpendicular to the sample surface may bring non-uniform distortions to different diffraction spots. As a consequence, the convolution of structural dynamics and such TEF is complex. To ultimately separate the TEF effect from structural dynamics in reflection UED, it is crucial to have a better understanding on the origin and evolution of the TEF and its influence on the probe electrons. Moreover, the TEFs induced by femtosecond pump laser pulses may influence the angular and energy resolution of photoelectrons in TR-ARPES studies [21-23, 30], which also leads to the necessity of its investigation.

Previously, studies on light induced electron emissions has been focused on the quantum yield or energy spectroscopy of photoelectrons [31-34], while the temporal evolution of TEFs is sparsely understood. Recently, ultrafast electron deflection and shadowgraph [10, 28, 35-37] have provided a direct monitor to transient electromagnetic fields, combining the intrinsic field sensitivity of electrons with the ultrahigh temporal resolution provided by a laser-pump electron-probe configuration.

In this contribution, the TEFs generated by femtosecond laser pulse irradiation of a 25-nm thick aluminum film have been investigated by picosecond electron deflection. Under laser intensities on the order of $10^{10}$ W/cm$^2$, it is shown that the TEFs at 120 μm above the metallic surface last more than one nanosecond with a maximum strength on the level of $10^4$ V/m. The experimental results were explained by a "three-layer" analytic model, which indicates that the observed TEFs were mainly attributed to the thermionic emission of electrons with an initial velocity of 1.4 μm/ps and a charge density of approximately $10^7$ e$^-$/mm$^2$. Based on the dynamics of the TEFs revealed in this study, we further evaluated their influence on UED and TR-ARPES.

**2. Ultrafast electron deflection configuration**

The laser-pump electron-probe experimental configuration, as shown in Figure 1, includes a Ti:sapphire laser system (1 kHz, 800 nm, 70 fs, 1 mJ/pulse ), a photoelectron gun driven by ultraviolet pulses, a magnetic lens, a sample holder attached to a 5-axial manipulator, an imaging system, and an ultrahigh vacuum chamber. The main laser beam was split into two parts: 90% was used as the pump and directed to a linear translation stage to precisely control its relative time difference (delay time) with respect to the probe beam. The pump beam was focused to a diameter of 0.8 mm (1/e$^2$) and normally impinged onto a freestanding 25 nm thick aluminum sample, which was a paradigm for UED experiments and prepared according to a routine procedure [38]. The pump intensities of interest was varied from 2.9 to 7.1×$10^{10}$ W/cm$^2$ (2~5 mJ/cm$^2$ fluence), in the same range as generally applied in time-resolved diffraction studies [38-40] and well below the ~10 mJ/cm$^2$ damage threshold of aluminum [41, 42]. Under the pump intensities used in this study, the temporal evolutions of the observed deflection angles are repeatable even after a large number of laser shots on the sample. We also inspected the sample after the experiments by an optical microscopy and no observable damage was found on its surface. The remaining 10% of the main beam was converted to 266 nm ultraviolet light through a frequency tripler and directed to the photocathode of the electron gun, a 30-nm silver layer coated on sapphire disc, to generate ultrashort electron pulses. The probe electrons were accelerated to 59 keV, collimated and focused to a diameter of ~200 μm ($1/e^2$) by the magnetic lens with their centroid at 120 μm above the sample surface. After passing through the sample area, the deflected probe electrons were recorded by the two dimensional imaging system containing a phosphor screen, a multi-channel plate (MCP) and a charge-coupled device (CCD) camera. Each electron deflection image was acquired with 1-s CCD exposure time to accumulate $10^3$ electron pulses and the signal-to-noise ratio was further improved by averaging more than 15 independent measurements of the electron deflection pattern at each delay time. The ~6 ps temporal resolution of the current setup is mainly limited by the travelling time of the 59 keV probe electrons through the laser-aluminum interaction area, which has a dimension similar to that of the pump beam. The time zero was defined as the onset of the observable deflections of the probe electrons.

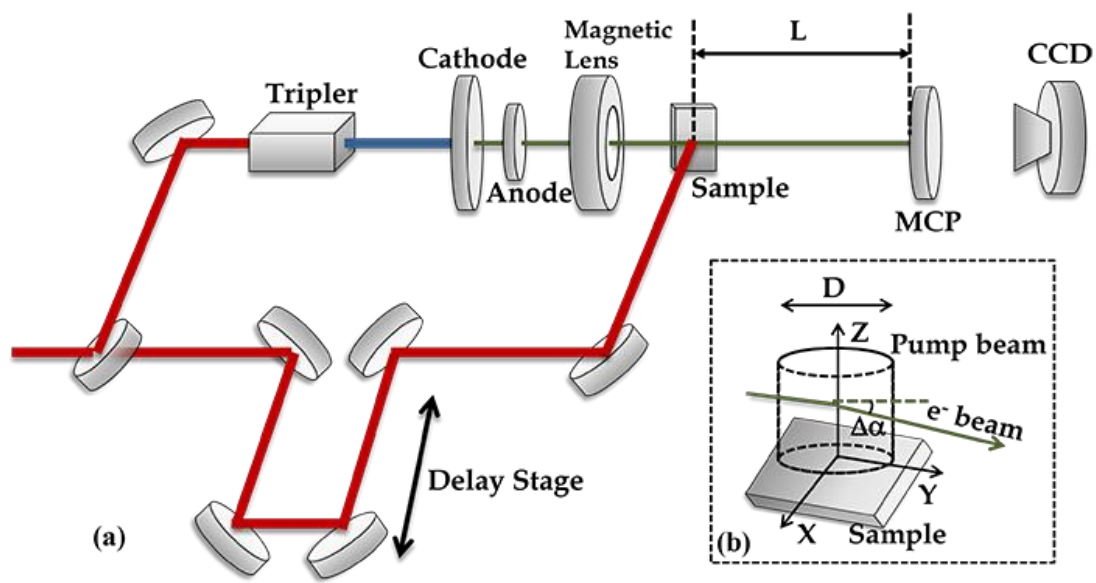

**Figure 1,** The schematic illustration of ultrafast electron deflection. (a), the experimental

configuration. (b), a detailed view of the sample area. Z axis denotes the sample surface normal direction, while X and Y axis are parallel to the sample surface. The probe electron beam travels along the Y axis before entering the TEF area and its centroid position at the Z axis is $Z_0 = 120$ μm. $\Delta\alpha$ is the deflection angle of the probe electron beam. The positive and negative deflection angles represent that the probe electron beam is deflected toward and away from the sample surface, respectively.

## 3. Electron deflection data analysis

The time-dependent evolution of the TEF was represented by the corresponding deflection angle of the probe electrons at each delay time. In order to calculate the electron deflection angle, we first integrated the 2D deflection pattern (see Figure 2(a)) along the X and Z directions to obtain two 1D intensity profiles, which were fitted by Gaussian functions to derive the peak positions as the centroid coordinates of the probe electrons. Then, the absolute change of the centroid position at each delay time, $\Delta R_i(t)$ ($i = x, z$), was obtained by subtracting the averaged value before time zero. Because the TEF is mainly perpendicular to the sample surface, only the 1D intensity distribution along the Z-axis is considered in the study presented here. Taking account of the distance between the sample and the detector ($L = 0.46\,m$) and the small angle approximation, the deflection angle of the probe electrons is $\Delta\alpha_z(t) \approx \arctan[\Delta\alpha_z(t)] = -\dfrac{\Delta R_z(t)}{L}$, where the minus sign represents that the positive direction of the TEF parallels the positive direction of Z-axis. The time-dependent evolution of the electron deflection angle actually represented the averaged TEF strength at 120 μm above the sample surface, where the centroid of the probe electrons locates. Meanwhile, the width and amplitude derived from the Gaussian fitting of the intensity profile along the Z-axis were normalized by their corresponding averaged values before time zero to deduce their time-dependent evolution.

## 4. Results and discussions:

### 4.1. The evolution of transient electric field

Upon femtosecond laser excitation of aluminum, the optical energy is rapidly deposited into conduction electrons because their heat capacity is several orders of magnitude smaller than that of the lattice [38]. Some energetic electrons overcome the ~ 3.9 eV work function of the nanosized thin aluminum [43], and escape from the sample surface via thermionic and/or multi-photon emission [1-5]. The evolution of the emitted electrons and the positive ion layer eventually determine the formation and decay of the TEF.

The TEF represented by the deflection of the probe electrons' centroid is the average strength sensed during their interrogation of the TEF. The relation between the electron deflection angle, $\Delta \alpha_z(t)$, and the averaged TEF strength detected by the probe electrons at each delay time, $\overline{E}_z(z_0, t)$, is described by the following equation:

$$\Delta \alpha_z(t) = -\frac{\Delta R_z(t)}{L} = -\frac{\Delta V_z(t)}{V_e} = \frac{\overline{E}_z(z_0,t)qD}{mV_e^2} = \frac{\overline{E}_z(z_0,t)qD}{m_e V_e^2 / \sqrt{1-V_e^2/c^2}} \quad (1)$$

where $V_e$ is the velocity of the probe electrons along their initial propagation direction, $\Delta V_z$ is the velocity change of the probe electrons along the sample surface normal direction induced by the TEF, $Z_0=120$ μm is the centroid position of the probe electrons along the Z direction before deflection, $D$ is the diameter of the pump laser focal spot at the sample position, $q$ and $m_e$ are the elementary charge and rest mass of an electron, and $m = m_e \left/ \sqrt{1 - V_e^2/c^2} \right.$ is the relativistic mass of an electron, respectively.

The time-dependent deflections of the probe electrons and the corresponding TEF strengths were shown in Figure 2. The typical error bar is mainly contributed by the pointing jitter of the probe electron beam, while the pointing jitter of the pump laser beam is negligible due to a much larger irradiation diameter. The maximum averaged electric field appears at t=158 ps and according to Eq. (1), its strength range from 32 to 53 kV/m with the pump intensities varying from 2.9 to 7.1×$10^{10}$ W/cm$^2$. However, the temporal evolution of the TEF remains the same under different pump intensities. The evolution of the TEF consists of three steps: (i). The negative deflection of the probe electrons reaches its minimum within 33 ps after laser irradiation. Because the centroid of the emitted electrons moves toward to that of the probe beam, the direction of the TEF at $Z_0=120$ μm is along the negative Z direction. Therefore, as shown in Figure 2(b), the probe electrons are deflected along the positive Z-axis by the TEF immediately after laser irradiation. Meanwhile, as shown in Figure 2(e), the deflection angle grows as a function of delay time and reaches its minimum at t=33 ps when the centroid of the probe electrons is probably deflected farthest away from the sample surface. (ii). The deflection of the probe electrons evolves from its minimum to zero at ~60 ps and reaches the positive maximum deflection at around 158 ps. Due to Coulomb repulsion inside the emitted electrons and the attractive force from the positive surface ion layer, a significant number of the emitted electrons decelerate from their initial emitting velocities and fall back into the sample while the remaining electrons effectively escape from the sample [44]. Given that the initial emitting velocity of the laser-excited electrons is on the order of 1 μm/ps [28], the "fallen back" electrons are always below the centroid position of the probe electrons and become the dominate contribution to the negative TEF strength at the centroid positon. Therefore, accompany with a large amount of the emitted electrons below $Z_0=120$ μm returning back to the sample, the magnitude of the deflection decreases after t=33 ps. At ~60 ps, the deflection angle of the probe electrons is zero, which means the direction of the averaged TEF at $Z_0=120$ μm will change from the negative to the positive Z direction. After 60 ps, the attractive force from the surface ion layer dominates the positive deflection of the probe electrons. Meanwhile, after the "effectively emitted" electrons pass the entire probing area, both the ion layer and the "effectively emitted" electrons contribute to the positive deflection of the probe electrons. Therefore, the probe electrons reach their positive maximum deflection at t=158 ps. (iii). The positive deflection of the probe electrons follows a decay process. Due to the expansion and moving away of the emitted electrons, and the accompany decreasing of the charge density, the TEF strength continues to descend for at least one nanosecond.

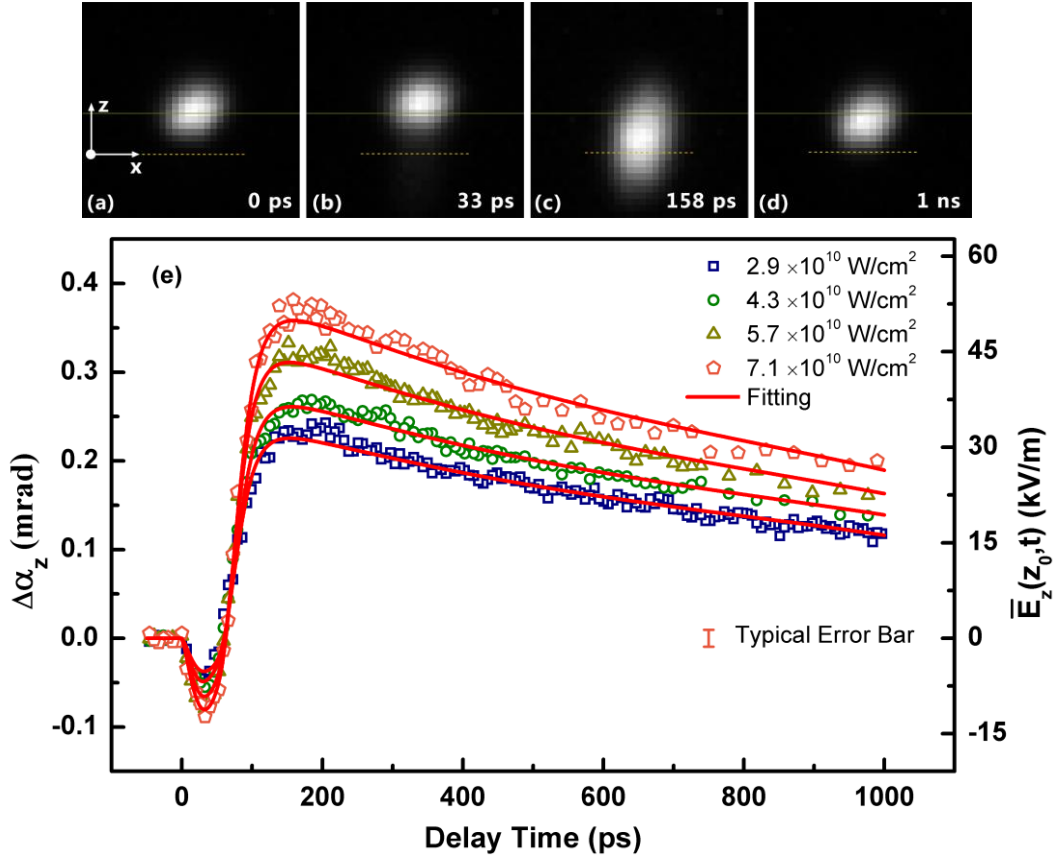

**Figure 2**, Deflection of the probe electron beam as a function of delay time. **(a)~(d):** Snapshots of probe electrons at a pump intensity of $7.1 \times 10^{10}$ W/cm$^2$. The horizontal dash and solid lines are marked for the sample position and the centroid of the probe electrons at the Z direction before deflection, respectively. **(e):** Time-dependent evolution of the electron deflection angles at various pump intensities. The negative and positive $\Delta\alpha_z$ indicates that the transient centroid of the probe electrons located above and below the horizontal solid line, respectively.

## 4.2 The evolution of the TEFs explained by a "three-layer" model

The evolution of the TEFs observed above results from the complex nonlinear many-body interactions between the emitted electrons and the positive surface charges, both evolving fast with time. In order to understand the origin of the observed TEFs and further estimate their influence on time-resolved electron scattering studies, we developed a "three-layer" analytical model. This model is aimed at reproducing the main features of our TEF measurements, together with an evaluation of the key parameters of the emitted electrons and the remaining positive ions. The "three-layer" model, which is illustrated in Figure 3, describes three types of charges that contribute to the TEFs: the positive surface charges that are due to the emission of electrons upon laser excitation, the emitted electrons that will return to the sample (fallen back electrons) and the emitted electrons that will effectively escape from the sample (effectively emitted electrons). The charge density and the longitudinal (perpendicular to the sample surface) and transversal (parallel to the sample surface) dimensions of these three types of charge layers evolve with time and determine the observed TEFs. The "three-layer" model is developed from the two-disk models reported before [28, 36, 45], which either neglected the thickness of the emitted electron layer or described the two kinds of electrons with different behaviors by a single Gaussian distribution. In the "three-layer" model, the total-emitted electrons are represented by the combination of two different Gaussian distributions, which correspond to the different behaviors of emitted electrons. It is expected to have an improved description of the actual behaviors of the total-emitted electrons.

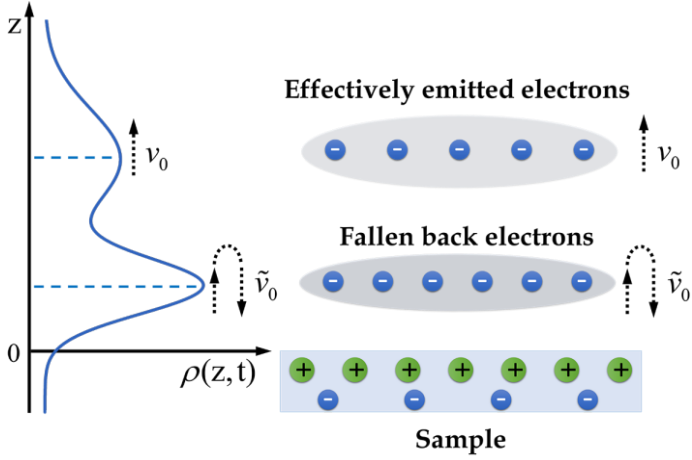

**Figure 3:** Schematic illustration of the "three-layer" model describing the evolution of emitted electrons and positive surface charges. The electron distribution function shown at the left side is for demonstration purpose only. In actual case, the effectively emitted electrons only account for a small portion of the total emitted electrons. Therefore, the effectively emitted electrons appears as a small shoulder in the distribution of the total-emitted electrons.

The initial transversal dimensions of the positive charges and the emitted electrons are assumed to be equal to the pump laser spot with a diameter of $D$. The positive charge layer contracts at an average speed $V_{w1}$ due to the degrading of mirror charge effects and neutralization. The emitted electrons expand at an average speed $V_{w2}$ and $V_{w3}$ due to the transversal and longitudinal Coulomb interaction, respectively. The longitudinal distribution of the positive charges is assumed to be a Delta function, which means that all the positive charges are confined at the surface of the sample, z=0 μm. The distribution function of the total-emitted electrons, $\rho(z,t)$, is defined by the following relation:

$$\rho(z,t) = (1-\alpha)\rho_E(z,t) + \alpha\rho_F(z,t) \qquad (2)$$

where $\alpha$ is the ratio of the "fallen back" electrons to the total-emitted electrons. $\rho_E(z,t)$ and $\rho_F(z,t)$ are two normalized Gaussian distributions representing the "effectively emitted" and the "fallen back" electrons along the Z direction (longitudinal), respectively. The time-dependent widths and peak positions of these two Gaussian functions have the same initial values at time zero. The "effectively emitted" electrons are assumed to move away from the sample surface with an initial emitting velocity $v_0$. The "fallen back" electrons decelerate with a rate $a$ from the same initial velocity $v_0$ to zero, then accelerate toward the sample surface with the same rate and eventually neutralize the positive ion layer. The propagating velocities are the center-of-mass (CoM) velocities of the effectively emitted" and the "fallen back" electrons. As a result of the velocity distribution inside the emitted electrons, the width of the electron distribution function also evolves with time. The symbols used in the model are summarized in Table 1.

**Table 1,** Symbols used in the "three-layer" model.

| Symbol | Description |
|---|---|
| $\rho(z,t)$ | Distribution function representing the total-emitted electrons |
| $\rho_E(z,t)$ | Gaussian distribution function representing the "effectively emitted" electrons |
| $\rho_F(z,t)$ | Gaussian distribution function representing the "fallen back" electrons |
| $\alpha$ | Ratio of the "fallen back" electrons to the total-emitted electrons |
| $a$ | Decelerating rate of the fallen back electrons |
| $v_0$ | Initial CoM velocity of the emitted electrons |
| $D$ | Initial transversal diameter of the ions and the emitted electrons (pump laser diameter) |
| $V_{w1}$ | Transversal contraction speed of the positive charges |
| $V_{w2}$ | Transversal expansion speed of the emitted electrons |
| $V_{w3}$ | Longitudinal expansion speed of the emitted electrons |
| $\sigma_0$ | Charge density at time zero |
| $\varepsilon_0$ | Dielectric constant of vacuum |

For simplicity, the averaged TEF strength was represented by the strength at the centroid position of the probe electrons, which is the contribution of both the positive surface ion layer and the negative electrons described below:

$$E_z(z_0,t) = \frac{\sigma_0}{2\varepsilon_0} \cdot \left\{ \left[1 - \frac{z_0}{\sqrt{z_0^2 + (D/2 - v_{w1} t)^2}}\right] \cdot \left[1 - \int_{-\infty}^{0} \rho(z,t) dz\right] \right.$$
$$- \int_{0}^{z_0} \rho(z,t) \cdot \left[1 - \frac{z_0 - z}{\sqrt{(z_0 - z)^2 + (D/2 + v_{w2} t)^2}}\right] dz \quad (3)$$
$$\left. + \int_{z_0}^{+\infty} \rho(z,t) \cdot \left[1 - \frac{z - z_0}{\sqrt{(z_0 - z)^2 + (D/2 + v_{w2} t)^2}}\right] dz \right\}$$

The experimental data were acquired from the shifting of the probe electrons centroid, which originally locates at 120 μm above the sample surface, therefore, the value of $z_0$ is 120 μm in Eq. (3). Together with Eq. (1), the deflections of the probe electrons at all four pump intensities were well fitted as depicted in Figure 2. Among all the seven fitting parameters ($\alpha, v_0, \sigma_0, a, V_{w1}, V_{w2}, V_{w3}$), the temporal evolutions of the transient electric field are most sensitive to $\alpha, v_0$ and $\sigma_0$, which are listed in Table 2 for detailed discussion. In addition, all fitting parameters were set as free variables with no constrains. Because the photon energy of the pump laser is 1.55 eV, much lower than the 3.9 eV work function of the nanosized thin aluminum [43], electrons are expected to be induced by thermionic and/or multi-photon emission instead of single-photon process. We performed the pump intensity dependence experiments to further distinguish these two emission mechanisms. According to Fowler-DuBridge theory [46] that described the electron emission from a solid surface, the electron yield of the n-th order photoemission is proportional to $I^n$, where $I$ is the pump laser intensity. However, the amount of the total-emitted electron charges is found to depend linearly on the pump intensity, as depicted in Figure 4. This linear relation indicated that thermionic emission is the dominant mechanism within the pump intensities applied here and the contribution of multiphoton emission is insignificant, which consists with the previous theoretical prediction [44].

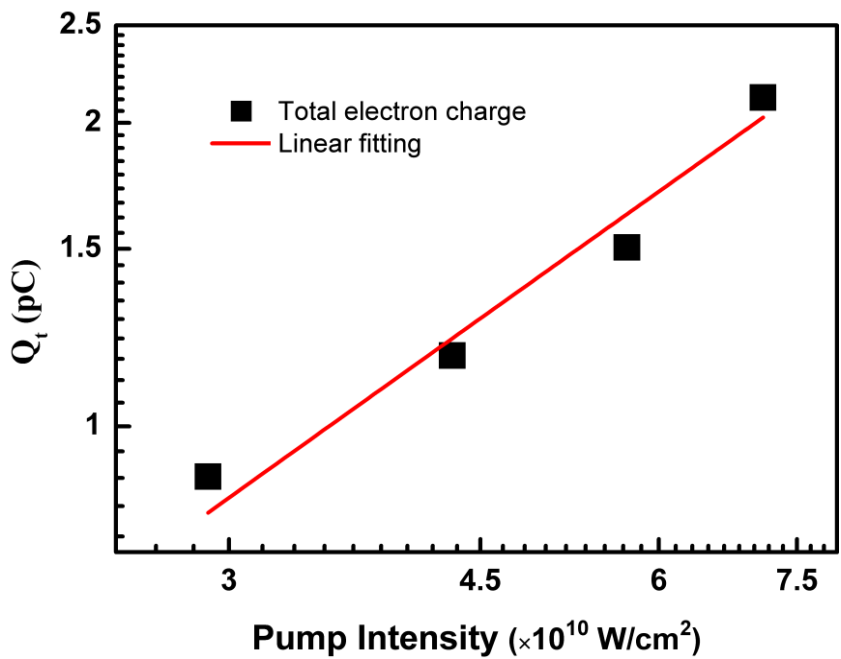

**Figure 4,** Linear dependence of the amount of the total-emitted electron charges on pump intensities.

**Table 2,** Key parameters obtained from the "three-layer" model. $Q_t$ and $Q_e$ represent the total-emitted and "effectively emitted" electron charges, respectively.

| Pump Intensity ($\times 10^{10}$ W/cm²) | $v_0$ (μm/ps) | $\sigma_0$ ($\times 10^7$ e/mm²) | $\alpha$ (%) | $Q_t$ (pC) | $Q_e$ (pC) |
|---|---|---|---|---|---|
| 2.9 | 1.44±0.003 | 1.1±0.05 | 75.5±1.2 | 0.88 | 0.27 |
| 4.3 | 1.42±0.003 | 1.5±0.04 | 78.4±0.7 | 1.21 | 0.32 |
| 5.7 | 1.44±0.003 | 1.9±0.06 | 80.4±0.7 | 1.53 | 0.37 |
| 7.1 | 1.40±0.002 | 2.6±0.05 | 84.0±0.3 | 2.09 | 0.42 |

The fitting results also suggest that, the temporal evolution of the electron distribution functions changes slightly with the increasing of the laser intensity, while the "fallen back" ratio $\alpha$ and the initial charge density $\sigma_0$ increase. Therefore, according to Eq. (2) and Eq. (3), $\alpha$ and $\sigma_0$ only modulate the amplitude of the charge distribution function and the strength of TEF, respectively. This agrees with the experimental results that, the temporal evolutions of the electron deflections are similar at all pump intensities increasing from 2.9 to 7.1$\times 10^{10}$ W/cm², while the deflection magnitude grows accordingly. The initial CoM emitting velocity of the electrons was fitted to be about 1.4 μm/ps. Therefore, the "effectively emitted" electrons travel ~220 μm and pass the entire probe electron beam at about 158 ps, which corresponds to the positive maximum deflection of the probe electrons depicted in Figure 2. As for the "fallen back" electrons, they are decelerated to zero velocity at ~30 μm above the sample and start to fall into the sample from this position. The strong Coulomb repulsion along the longitudinal direction causes the return of more than 75% of the total-emitted electrons back to the sample together with the attraction force from the positive surface charges.

**4.3. Influence of TEF on time-resolved electron scattering studies.**

Pump-probe technique is the primary tool for the studies of transient phenomenon, especially on the picosecond to femtosecond time scale. The pump and probe sources can vary from different combinations of optical, THz, X-ray, and electron pulses. Among all these combinations, ultrafast electron diffraction, which is based on laser-pump electron-probe, has been an effective method that provides direct access to structure evolution with atomic spatial-temporal resolutions. In UED studies, structural dynamics are generally obtained from the time-dependent evolution of electron diffraction angles, line widths, and integrated intensities. However, both structural dynamics and transient electric fields can affect the behavior of the probe electron beam, which gives rise to difficulties in the interpretation of electron diffraction data [25, 28, 45].

In this study, the TEF at 120 μm above the metallic surface was found to be on the order of $10^4$ kV/m under moderate pump intensities, which are generally applied in UED studies. Its influences on the deflection angle, width, and peak intensity of the probe electron beam profile are depicted in Figure 5 and some characteristic parameters are given in Table 3. For the first tens of picoseconds after laser irradiation, the deflection angle of the probe electrons is on the order of tens micro radians, which is comparable to the typical changes of the diffraction angle induced by structural dynamics in reflection UED [24-26]. With the fitting parameters obtained under the pump fluence of $7.1\times10^{10}$ W/cm$^2$, we further calculated the TEF gradient along the Z direction according to the "three-layer" model and evaluated its influence on the broadening of the probe beam profile. The results imply that, the maximum beam width reached at t=92 ps is 2.2 times of that before time zero, which is in good agreement with the 2.3-times experimental value. This good agreement suggests that, applying Eq.(3) for the averaged TEF is reasonable and the "three-layer" model is self-consistent. In addition, the broadening of the beam width also induces the attenuation of the peak intensity. Both the width and peak intensity recover toward their original values before time zero along with the decay of the TEF.

In general, the results presented here indicate that, TEFs widely exist in UED studies, and in a reflection configuration, it can affect the position, width and peak intensities of the probe electron beam profile, which may cause misinterpretations to the structural dynamics extracted from diffraction patterns. In future UED studies, the pump laser induced TEFs should be evaluated *in situ* for a closer understanding of the structural dynamics and a better resolution. Meanwhile, the strong electric field above the sample surface may also contribute to the transient structure change, which has not been considered in the previous UED studies. Therefore, further efforts are necessary to access the role of TEF effects on transient structure changes.

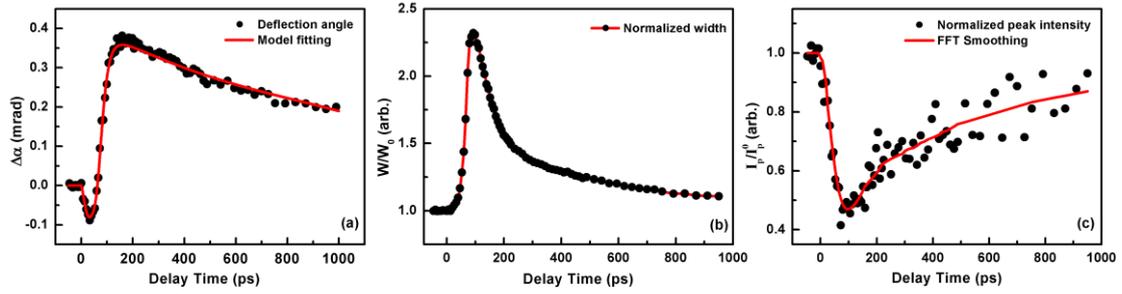

**Figure 5:** Time-dependent evolution of the **(a)** deflection angle, **(b)** width, and **(c)** peak intensity of the probe electron beam at a pump intensity of $7.1\times 10^{10}$ W/cm$^2$. The evolution of the deflection angle was fitted by the "three-layer" model. The peak intensity was normalized to the value before time zero and smoothed with Fast-Fourier-Transformation filtering.

**Table 3:** Typical parameters of the dynamical position presented in the deflection angle, width and peak intensity of the probe beam profile.

|  | $T_{max}$ (ps) | Maximum Value |
| --- | --- | --- |
| $\Delta\alpha$ | 158 | 0.38 mrad |
| $W/W^0$ | 92 | 2.3 |
| $I_p/I_p^0$ | 92 | 0.47 |

The TEF effects may also be an important issue in TR-APERS studies emerged recently, which provide a temporal, angular and energy resolution of photoelectrons on the order of sub-picoseconds, tenth of a degree and milli-electronvolts, respectively[23, 47]. In these TR-ARPES studies, samples are pumped by a femtosecond laser pulse and probed by ultraviolet (UV) photons to reveal the time-dependent photoemission spectroscopy at the first few picoseconds after laser excitation. It is generally believed that these photoelectrons is induced by the UV photons and the space charge effect of such electrons has been extensively studied [32, 48, 49] to improve the energy resolution. However, even at a low fluence/intensity, the femtosecond pump laser pulse can rapidly heat up the electron system and may generate photoelectrons. This additional effect, which could also be an influencing factor to the energy resolution of TR-ARPES, has rarely been assessed [50].

We estimated the averaged electric field strength on aluminum surface by the "three-layer" model under the lowest pump intensity used here. As presented in Figure 6, it indicated that the averaged electric field strength within several micrometers above the sample surface is on the order of 100 kV/m for the first few picoseconds. Its modulation to the photoelectrons is on the order of $10^2$ meV, which may influence the understanding of TR-ARPES results and limit the improvement of the energy resolution to better than milli-electronvolts. Although the samples of interest in TR-ARPES studies are mainly superconductors or topological insulators, the study presented here may bring into attention that, it is insufficient to only consider the effect of UV induced photoelectrons, and the TEF induced by the pump laser pulse should also be evaluated to optimize the performance of TR-ARPES.

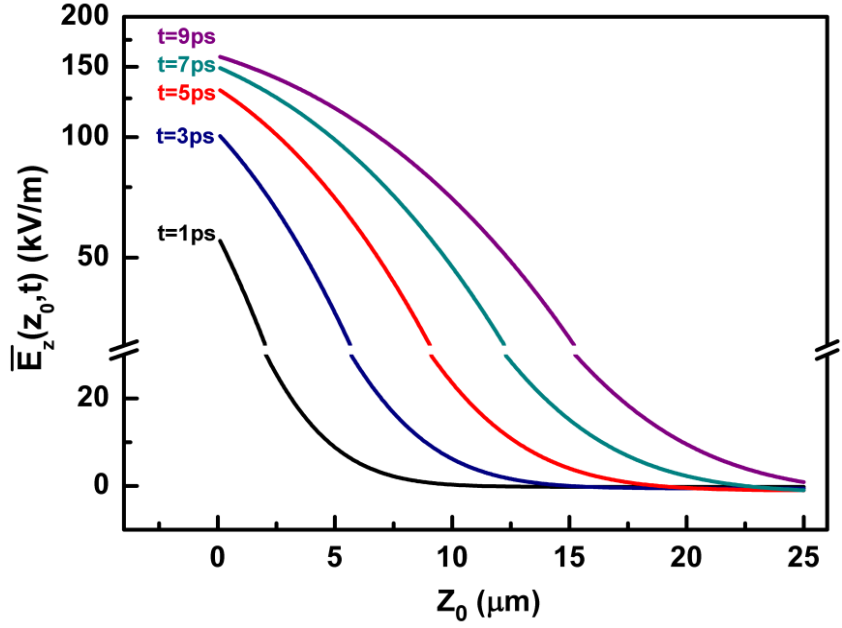

**Figure 6,** The TEF strength along the Z direction for the first few picoseconds, which is predicted by the "three-layer" model and the fitting parameters under the lowest pump intensity, $2.9 \times 10^{10}$ W/cm$^2$.

However, limited by the experimental configuration of this study and the simplified "three-layer" analytical model, we can only obtain some finite insights into the transient electric field on the metallic surface. Inspired by proton radiography [51], in our further efforts, we will experimentally investigate the spatial-temporal evolution of the TEFs by ultrafast electron radiography. A better understanding of the TEFs may help to improve the resolution and accuracy of time-resolved studies that involved with electrons.

## 5. Conclusion:

We used ultrashort electron pulse to directly monitor the femtosecond laser induced transient electric field above the metal surface, which was found to build up in hundreds of picoseconds and decay within nanoseconds. Its strength is on the scale of $10^4$ V/m at 120 μm above the sample surface under the pump intensities on the order of $10^{10}$ W/cm$^2$. The experimental results were explained by a "three-layer" analytic model, and the observed TEFs were attributed to the thermionic emission of electrons with an initial velocity of ~1.4 μm/ps and a charge density of approximately $10^7$ e$^-$/mm$^2$. The study presented here also indicate that, besides deflection, the probe electron beam width, peak intensity and energy dispersion can also been modulated by the transient electric field. Therefore, for time-resolved electron scattering studies, such as ultrafast electron diffraction and time-resolved angle-resolved photoemission spectroscopy, the transient surface electric field should be considered and evaluated *in situ* for improved resolution and accuracy.

**Acknowledgements**

The authors would like to thank Dr. Dong Qian for helpful discussions on TR-ARPES, Mr. Xing Lin for aluminum deposition, and Mrs. Xin-Qiu Guo and Mr. Gang-Sheng Tong for Transmission Electron Microscopy inspection of the sample. This work was supported by the National Basic Research Program of China under Grant No. 2013CBA01500 and the National Natural Science Foundation of China under Grants No. 11004132, 11121504 and 11327902. J. Cao would like to acknowledge the support from Nation Science Foundation under Grant No. 1207252.